# Modeling of an improved positive corona thruster and actuator

Alexandre A. Martins

Institute for Plasmas and Nuclear Fusion & Instituto Superior Técnico,
Av. Rovisco Pais, 1049-001 Lisboa, Portugal
(aam@ist.utl.pt)

**Abstract**

In this work we are going to perform a simulation of a positive corona discharge in nitrogen gas, using two different asymmetric capacitor geometries. This work represents an improvement over previous work in order to accomplish higher ion wind velocities as well as stronger electrostatic propulsion forces on all the involved electrodes. In our model, the used positive ion source is a small diameter wire, which generates a positive corona discharge in nitrogen gas directed to the ground (or negatively charged) electrodes. By applying the fluid dynamic and electrostatic theories, all hydrodynamic and electrostatic forces that act on the considered geometries will be computed in an attempt to demonstrate the greater performance of the new developed geometries.

## 1. Introduction

Direct current/voltage corona discharges are very useful to produce meaningful propulsion forces and ion wind velocities that, like plasma actuators, could also be applied for many different practical purposes. They may be useful as a wind blower in order to substitute mechanically rotating small wind fans or even as propulsion units for small UAV's (unmanned aerial vehicles). Other possible uses are the improvement of the efficiency of turbo-machinery by reducing turbulence, or also in aerodynamic boundary layer control, aircraft drag reduction or wing lift increase, noise reduction or other internal flow configurations, with the advantage of using no moving mechanical parts [1-10].

Our present work has the purpose to improve on the performance of previous high performance geometries [11, 12] with an aim to increase the generated ion wind velocity as well the propulsion force on simple asymmetric capacitor setups. This greater performance will broaden the practical applicability and use of such systems.

We are going to investigate two different geometries implementing a numerical simulation in nitrogen gas (as a first approximation to air), at atmospheric pressure, in order to obtain the resultant electro-hydrodynamic (EHD) flow and forces when a positive corona discharge is established. The first geometry uses two adjacent grounded electrode spheres of 0.5 cm in diameter. This structure is an asymmetric capacitor with only one corona wire centered at (-0.03 m, 0 m) distanced approximately 3 cm from the two grounded spheres which are centered at (0 m, -0.01 m) and (0 m, 0.01 m), and that are distanced between themselves by 1.5 cm, as can be seen in figure 1.a) (the space between each dot represents 1 cm). The positive corona wire has a radius of 0.25 μm, which is much smaller than the radius of 0.25 cm of the grounded spheres.

In the second geometry we add two cathode spheres also with 0.5 cm in diameter (figure 1.b)) that are centered at (0.01 m, -0.01 m) and (0.01 m, 0.01 m), and are distanced from the ground spheres by 0.5 cm (surface to surface). Since we are going to apply a tension of -15 kV to the extra cathode spheres, we have chosen a distance of 0.5 cm between the ground and cathode spheres in order to have an electric field slightly below the atmospheric discharge limit of 33.1 kV/cm. The usefulness of adding the cathode spheres will be that the positive ions emitted by the corona wire will be further accelerated in the acceleration channel between the ground and cathode spheres, generating more momentum in the gas and thrust in the electrodes.

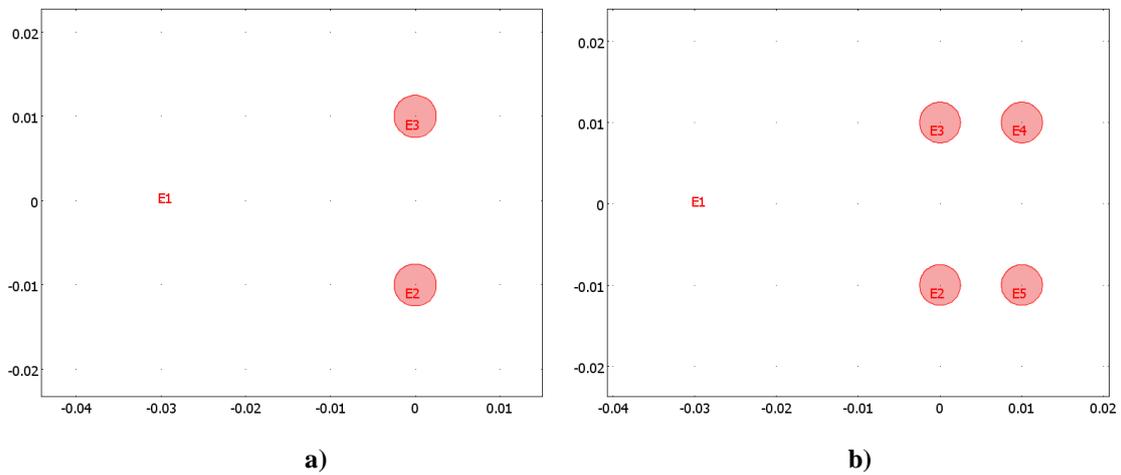

**Figure 1. a)** Asymmetric capacitor with the corona wire (E1) at (- 0.03 m, 0 m) and two grounded spheres (E2 and E3) 0.5 cm diameter and centered respectively at (0 m, -0.01 m) and (0 m, 0.01 m); **b)** two extra cathode spheres (E4 and E5) with 0.5 cm diameter centered at (0.01 m, -0.01 m) and (0.01 m, 0.01 m).

## 2. Numeric Model

In our numerical model we will consider electrostatic force interactions between the electrodes and the ion space cloud, the moment exchange between the mechanical setup and the induced opposite direction nitrogen flow (EHD flow), nitrogen pressure forces on the structure and viscous drag forces. EHD flow is the flow of neutral particles caused by the drifting of ions in an electric field. In our case these ions are generated by a positive high voltage corona discharge in the high curvature (the higher the radius of a sphere, the less is its curvature) portions of the electrodes. The corona wire has a uniform high curvature and therefore will generate ions uniformly [13]. The positively ionized gas molecules will travel from the corona wire ion source towards the collector (ground) colliding with neutral molecules in the process. These collisions will impart momentum to the neutral atoms that will move towards the collector as a result. The momentum gained by the neutral gas is exactly equal to the momentum gained by the positive ions accelerated through the electric field between the electrodes, and lost in inelastic collisions to the neutral molecules.

Corona discharges are non-equilibrium plasmas with an extremely low degree of ionization (roughly $10^{-8}$ %). There exist two zones with different properties, the ionization zone and the drift zone [14]. The energy from the electric field is mainly absorbed by electrons in the ionization zone, immediately close to the corona electrode, and transmitted to the neutral gas molecules by inelastic collisions producing electron-positive ion pairs, where the net space charge density $\rho_q$ will remain approximately zero ($\rho_q = 0$). However, the local volume space charge, in the drift zone, will be positive for a positive corona (and therefore $\rho_q = eZ_i n_i$ in the drift zone; $e$ is absolute value of the electron charge, $Z_i$ is the charge of the positive ionic species, $n_i$ is the positive ion density of $N_2^+$) because of the much higher mobility of electrons relative to the positive ions, and because the only region were (positive) ions will be generated is in the ionization region where the electrons have enough energy to accomplish that due to the much higher electric field intensity. Also, in the drift region the electrons and ions do not have enough energy to react with neutrals and also have too low density to react with other ionized particles [14]. Therefore, we will only consider nitrogen ($N_2^+$) positive ions in the drift zone, which is the most relevant nitrogen ion in atmospheric processes [13].

The ionic mobility ($\mu_i$) is defined as the velocity $v$ attained by an ion moving through a gas under unit electric field E (m$^2$ s$^{-1}$ V$^{-1}$), i.e., it is the ratio of the ion drift velocity to the electric field strength:

$$\mu_i = \frac{v}{E}. \tag{1}$$

The mobility is usually a function of the reduced electric field $E/N$ and $T$, where $E$ is the field strength, $N$ is the Loschmidt constant (number of molecules m$^{-3}$ at s.t.p.), and $T$ is the temperature. The unit of $E/N$ is the Townsend (Td), 1 Td = $10^{-21}$ V m$^2$. Since we are applying 28000 V to the corona wire across a gap of 3 cm towards the ground electrode, the reduced electric field will be approximately 38 Td or 38 x $10^{-17}$ Vcm$^2$ (considering that the gas density N at 1 atm, with a gas temperature $T_g$ of 300 K is N=2.447 x $10^{19}$ cm$^{-3}$). According to Moseley [15], the mobility $\mu_i$ of an ion is defined by:

$$\mu_i = \mu_{i0}(760/p)(T/273.16), \tag{2}$$

where $\mu_{i0}$ is the reduced mobility, $p$ is the gas pressure in Torr (1 atm = 760 Torr) and $T$ is the gas temperature in Kelvin. For our experimental condition of E/N = 38 Td, Moseley's measurements indicate a $\mu_{i0}$ of 1.83 cm$^2$/(Vs). Thus, at our operating temperature of 300 K, the mobility $\mu_i$ will be 2.01 cm$^2$/(Vs) or 2.01 x $10^{-4}$ m$^2$/(Vs).

Since the reduced electric field is relatively low, the ion diffusion coefficient $D_i$ can be approximated by the Einstein relation:

$$D_i = \mu_i \left(\frac{k_B T}{e}\right), \tag{3}$$

where $k_B$ is the Boltzmann constant. This equation provides a diffusion coefficient of $5.19 \times 10^{-6}$ m$^2$/s for our conditions.

The governing equations for EHD flow in an electrostatic fluid accelerator (EFA) are already known [16, 17] and described next; these will be applied to the drift zone only. The electric field **E** is given by:

$$\mathbf{E} = -\nabla V. \tag{4}$$

Since $\nabla \cdot \mathbf{E} = \dfrac{\rho_q}{\varepsilon_0}$ (Gauss's law), the electric potential **V** is obtained by solving the Poisson equation:

$$\nabla^2 V = -\frac{\rho_q}{\varepsilon_0} = -\frac{e(Z_i n_i - n_e)}{\varepsilon_0}, \tag{5}$$

where $n_e$ is the electron density (we are not considering negative ions) and $\varepsilon_0$ is the permittivity of free space. The total volume ionic current density $\mathbf{J}_i$ created by the space charge drift is given by:

$$\mathbf{J}_i = \rho_q \mu_i \mathbf{E} + \rho_q \mathbf{u} - D_i \nabla \rho_q, \tag{6}$$

where $\mu_i$ is the mobility of ions in the nitrogen gas subject to an electric field, *u* is the gas (nitrogen neutrals) velocity and $D_i$ is the ion diffusion coefficient. The current density satisfies the charge conservation (continuity) equation:

$$\frac{\partial \rho_q}{\partial t} + \nabla \cdot \mathbf{J}_i = 0. \tag{7}$$

But, since we are studying a DC problem, in steady state conditions we have:

$$\nabla \cdot \mathbf{J}_i = 0. \tag{8}$$

The hydrodynamic mass continuity equation for the nitrogen neutrals is given by:

$$\frac{\partial \rho_f}{\partial t} + \nabla \cdot (\rho_f \mathbf{u}) = 0 \tag{9}$$

If the nitrogen fluid density $\rho_f$ is constant, like in incompressible fluids, then it reduces to:

$$\nabla \cdot \mathbf{u} = 0. \tag{10}$$

In this case, the nitrogen is incompressible and it must satisfy the Navier-Stokes equation:

$$\rho_f \left( \frac{\partial \mathbf{u}}{\partial t} + (\mathbf{u} \cdot \nabla)\mathbf{u} \right) = -\nabla p + \mu \nabla^2 \mathbf{u} + \mathbf{f}. \tag{11}$$

The term on the left is considered to be that of inertia, where the first term in brackets is the unsteady acceleration, the second term is the convective acceleration and $\rho_f$ is the density of the hydrodynamic fluid - nitrogen in our case. On the right, the first term is the pressure gradient, the second is the viscosity ($\mu$) force and the third is ascribed to any other external force $\mathbf{f}$ on the fluid. Since the discharge is DC, the electrical force density on the nitrogen ions that is transferred to the neutral gas is $\mathbf{f}^{EM} = \rho_q \mathbf{E} = -\rho_q \nabla V$. If we insert the current density definition (Equation (6)) into the current continuity (Equation (8)), we obtain the charge transport equation:

$$\nabla \cdot \mathbf{J}_i = \nabla \cdot (\rho_q \mu_i \mathbf{E} + \rho_q \mathbf{u} - D_i \nabla \rho_q) = 0 \tag{12}$$

Since the fluid is incompressible ($\nabla \cdot \mathbf{u} = 0$) this reduces to:

$$\nabla \cdot (\rho_q \mu_i \mathbf{E} - D_i \nabla \rho_q) + \mathbf{u} \nabla \rho_q = 0 \tag{13}$$

In our simulation we will consider all terms present in Equation (13), although it is known that the conduction term (first to the left) is preponderant over the other two (diffusion and convection), since generally the gas velocity is two orders of magnitude smaller than the velocity of ions. Usually, the expression for the current density (Equation (6)) is simplified as:

$$\mathbf{J}_i = \rho_q \mu_i \mathbf{E}, \tag{14}$$

Then, if we insert Equation (14) into Equation (8), expand the divergence and use Equation (4) and Gauss's law we obtain the following (known) equation that describes the evolution of the charge density in the drift zone:

$$\nabla \rho_q \cdot \nabla V - \frac{\rho_q^2}{\varepsilon_0} = 0. \tag{15}$$

In Table I we can see the values of the parameters used for the simulation. We will consider in our model that the ionization region has zero thickness, as suggested by Morrow [18]. The following equations will be applied to the ionization zone in order to determine the proper boundary condition that we would have in the boundary between the ionization and drift zones and apply that directly on the surface of the corona wire, so that we take into account any ionization zone effects in our model. For the formulation of the proper boundary conditions for the external surface of the space charge density we will use the Kaptsov hypothesis [19] which states that below corona initiation the electric field and ionization radius will increase in direct proportion to the applied voltage, but will be maintained at a constant value after the corona is initiated.

In our case, a positive space charge $\rho_q$ is generated by the corona wire and drifts towards the ground electrode through the gap (drift zone) between both electrodes and is accelerated by the local electric field. When the radius of the corona wire is much smaller than the gap, then the ionization zone around the corona wire is uniform. In a positive corona, Peek's empirical formula [20-24] in air gives the electric field strength $E_p$ (V/m) at the surface of an ideally smooth cylindrical wire corona electrode of radius $r_c$:

$$E_p = E_0 \cdot \delta \cdot \varepsilon \left(1 + 0.308/\sqrt{\delta \cdot r_c}\right) \qquad (16)$$

Where $E_0 = 3.31 \times 10^6 V/m$ is the dielectric breakdown strength of air (we used the nitrogen breakdown strength, which is 1.15 times higher than that for air [21]), $\delta$ is the relative atmospheric density factor, given by $\delta = 298p/T$, where $T$ is the gas temperature in Kelvin and $p$ is the gas pressure in atmospheres ($T$=300K and $p$=1atm in our model); $\varepsilon$ is the dimensionless surface roughness of the electrode ($\varepsilon = 1$ for a smooth surface) and $r_c$ is given in centimeters. At the boundary between the ionization and drifting zones the electric field strength is equal to $E_0$ according to the Kaptsov assumption. This formula (Peek's law) determines the threshold strength of the electric field to start the corona discharge at the corona wire. Surface charge density will then be calculated by specifying the applied electric potential $V$ and assuming the electric field $E_p$ at the surface of the corona wire. The assumption that the electric field strength at the wire is equal to $E_p$ is justified and discussed by Morrow [18]. Although $E_p$ remains constant after corona initiation, the space charge current $\mathbf{J_i}$ will increase with the applied potential $V_c$ in order to keep the electric field at the surface of the corona electrode at the same Peek's value, leading to the increase of the surrounding space charge density and respective radial drift.

Atten, Adamiak and Atrazhev [23], have compared Peek's empirical formula with other methods including the direct Townsend ionization criterion and despite some differences in the electric field, they concluded that the total corona current differs only slightly for small corona currents (below 6 kV). For voltages above 6 kV (corresponding to higher space currents) the difference is smaller than 10% in the worst case, according to them.

For relatively low space charge density in DC coronas, the electric field $E(r)$ in the plasma (ionization zone) has the form [13]:

$$E(r) = \frac{E_p r_c}{r}, \qquad (17)$$

where $r$ is the radial position from the center of the corona wire. Since the electric field $E_0$ establishes the frontier to the drift zone, using this formula we can calculate the radius of the ionization zone ($r_i$), which gives:

$$r_i = \frac{E_p r_c}{E_0} = r_c \cdot \delta \cdot \varepsilon \left(1 + 0.308/\sqrt{\delta \cdot r_c}\right). \qquad (18)$$

Since we have chosen in our simulation for $r_c$ to be 0.025 mm, then $r_i$ would be 0.074 mm. Now we can calculate the voltage ($V_i$) at the boundary of the ionization zone by integrating the electric field between $r_c$ and $r_i$:

$$V_i = V_c - E_p r_c \ln(E_p/E_0), \qquad (19)$$

where $V_c$ is the voltage applied to the corona electrode and $r_c$ is in meters. This equation is valid only for the ionization zone. In our case it determines that if we apply 28000 Volts to the corona wire, then the voltage present at the boundary of the ionization zone becomes 26658.73 Volts, which is the voltage we apply directly to the surface of the corona wire in our simulation.

For the drift zone, Poisson equation (Equation (5)) should be used together with the charge transport equation (Equation (13)) in order to obtain steady state field and charge density distributions. The values of the relevant parameters for the simulation are detailed in Table 1.

**Table 1.** Value of parameters used for the simulation.

| Parameters | Value |
|---|---|
| Nitrogen density (T=300K, p=1atm), $\rho_N$ | 1.165 kg/m$^3$ |
| Dynamic viscosity of nitrogen (T=300K, p=1atm), $\mu_N$ | 1.775 x 10$^{-5}$ Ns/m$^2$ |
| Nitrogen relative dielectric permittivity, $\varepsilon_r$ | 1 |
| $N_2^+$ mobility coefficient, $\mu_i$ (for E/N = 38 Td) | 2.01 x 10$^{-4}$ m$^2$/(Vs) |
| $N_2^+$ diffusion coefficient, $D_i$ (for E/N = 38 Td) | 5.19 x 10$^{-6}$ m$^2$/s |
| Corona wire radius, $r_c$ | 25 μm |
| Ground electrode diameter | 0.5 cm |
| Distance between grounds / cathode spheres | 1.5 cm |
| Distance between ground and cathode spheres | 0.5 cm |
| Distance between wire and ground plane | 3.0 cm |
| Corona wire voltage, $V_c$ | 28000 V |
| Ground electrode voltage, $V_g$ | 0 V |
| Cathode electrode voltage | -15000 V |

Three application modes of the COMSOL 3.5 Multiphysics software are used. The steady state incompressible Navier-Stokes mode is used to resolve the fluid dynamic equations. The electrostatics mode is used to resolve the electric potential distribution and the electrostatic forces to which the electrodes are subjected. The PDE (coefficient form) mode is used to resolve the charge transport equation (Equation (13)). The parameters used for the simulation are shown in Table 1. The typical mesh of the solution domain consists of 53438 elements in a square of 0.2 m by 0.2 m as shown on figure 2.

Dirichlet boundary conditions were used in the PDE (coefficient form) module, where in the corona wire element an initial ion concentration of $1.6 \times 10^{-2}$ [Cm$^{-3}$] was used for the first geometry and an initial ion concentration of $2.2 \times 10^{-2}$ [Cm$^{-3}$] was used for the second geometry, whereas a zero ion concentration was used on the ground electrodes and on the domain frontiers. In the Incompressible Navier-Stokes module, an open boundary condition was implemented on the domain frontiers, and a wall (no slip) condition was implemented on all electrodes. In the Electrostatics module a zero charge/symmetry boundary condition was implemented on the domain frontiers, a Ground potential on the ground electrodes, -15000 Volts on the cathode spheres, and an electric potential of 28000 Volts on the corona wire, which according to Equation (19) transforms to 26658.73 Volts.

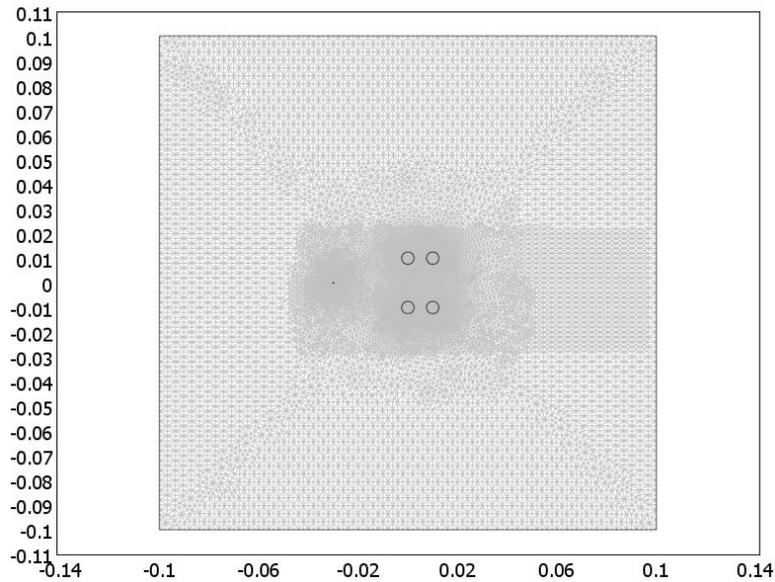

**Figure 2.** Typical mesh of the solution domain (0.2 m × 0.2 m) containing 53438 elements (units in m).

## 3. Numerical Simulation Results

In previous similar simulation results using Comsol 3.5 we have demonstrated that our theoretical results for the force on the electrodes in nitrogen had a 12% difference in relation to experimental measurements made in air [11], where an equal reduced electric field was used. On the other hand we have also demonstrated a stability or convergence up to the third decimal place above 30000 mesh elements [12]. In the current setup the corona wire generates a positive charge cloud which is accelerated through the gap towards the facing ground spheres and cathode electrodes. The interaction between all electrodes and the positive charge cloud will accelerate the nitrogen positive ions towards the grounded spheres and the ions will transmit their momentum to the neutral nitrogen particles by a collision process.

The neutral nitrogen will move from the positive corona wire to the grounded and negatively charged spheres, and the nitrogen neutral wind profile generated by the collisions with the accelerated ion cloud is shown in figure 3 for both geometries, were the top velocity achieved is 3.898 m/s for the first geometry and 4.895 m/s for the second geometry. Therefore the addition of the cathode spheres managed to increase the neutral wind velocity by 1 m/s.

The electric field vector distribution around the asymmetric capacitor (in the presence of the positive ion space charge) is shown in figure 4, where we can clearly see the directional electric field that accelerates the nitrogen positive ions generating the ion wind. The corresponding ion density profile is represented in figures 5 and 6, respectively for the first and second geometries.

The electrostatic force vectors for the electrodes in both geometries are shown in figure 7. The respective electrostatic force values along the horizontal (x axis) on the electrodes are given in tables 2 and 3.

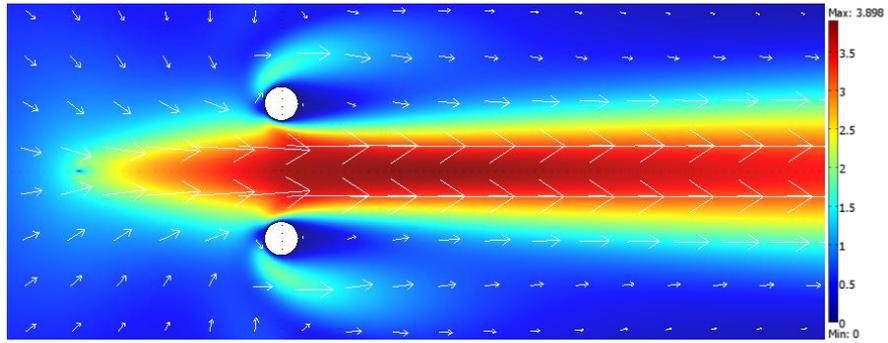

a)

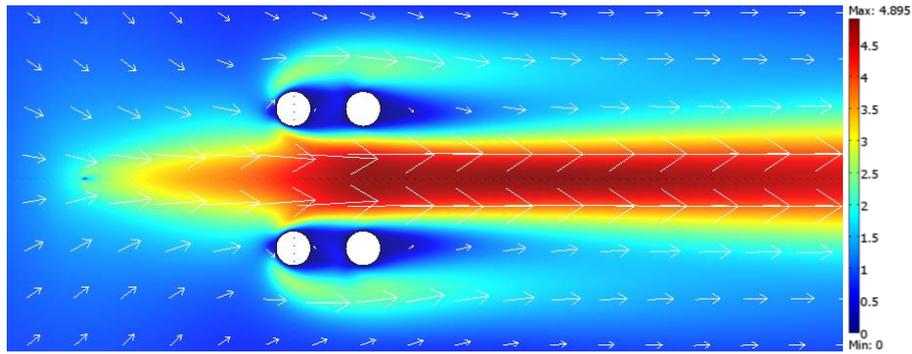

b)

**Figure 3.** Nitrogen velocity as surface map with units in m/s with proportional vector arrows, **a)** on first geometry, **b)** on second geometry.

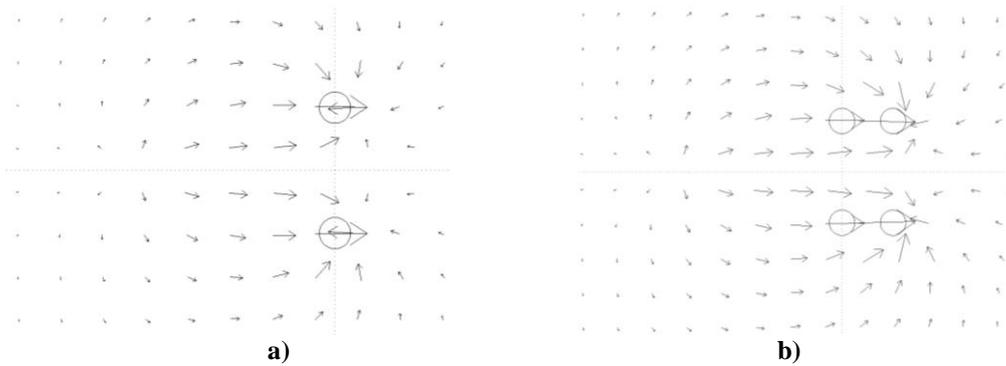

a)                                             b)

**Figure 4.** Distribution of the electric field vectors (arrows) in the asymmetric capacitor when the corona discharge is functioning, **a)** on first geometry, **b)** on second geometry.

The corona wire and nitrogen positive space charge will induce opposite charges in the ground and negative electrodes, which will be subjected to a strong electrostatic force. The results of the simulation show that the electrostatic forces $F_{ex}$ on the electrodes are the only relevant forces to consider, constituting 99.85% of the total force for the first geometry and 99.82% for the second geometry. The total hydrodynamic force $F_{HTx}$ (sum of pressure and viscosity forces) is very small because the pressure $F_{px}$ and viscosity $F_{vx}$ forces do not contribute in a relevant way in the present conditions. The total resultant force $F_{Tx}$ (sum of hydrodynamic and electrostatic forces) that acts on the first and second geometries is, respectively -0.2268 N/m and -0.3559 N/m (directed from the spheres to the wire).

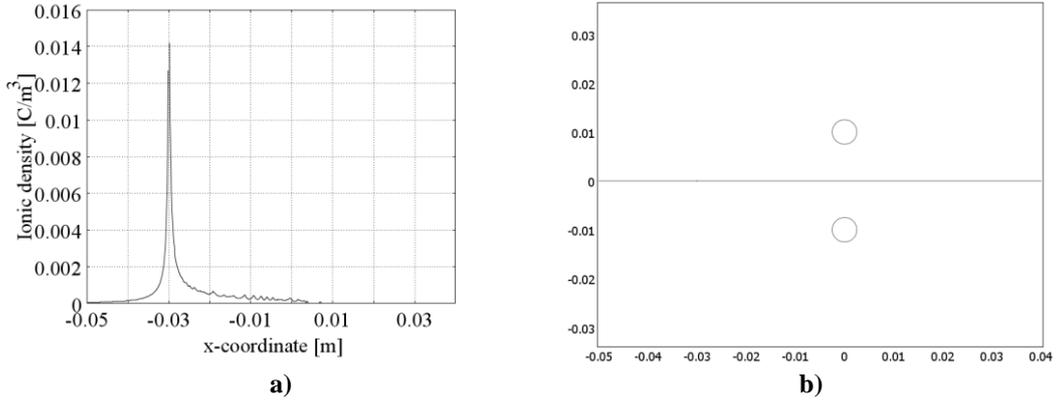

**Figure 5. a)** Ionic density distribution [C/m$^3$] for the first geometry, from (-0.05 m, 0 m) to (0.04 m, 0 m), **b)** horizontal line showing where the ion density is taken (units in meters).

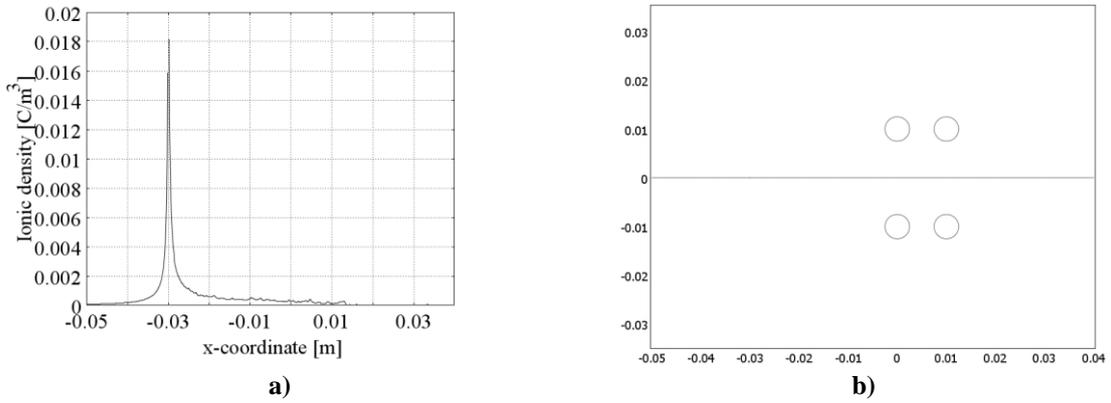

**Figure 6. a)** Ionic density distribution [C/m$^3$] for the second geometry, from (-0.05 m, 0 m) to (0.04 m, 0 m), **b)** horizontal line showing where the ion density is taken (units in meters).

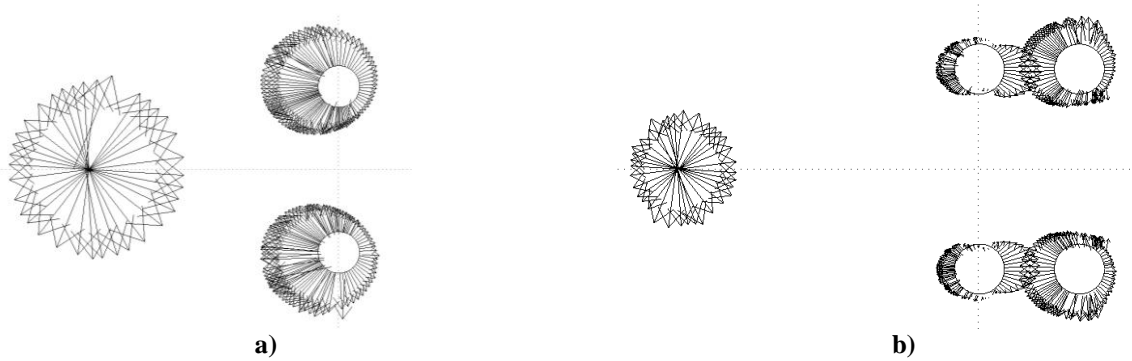

**Figure 7.** Electrostatic force with true relative magnitude between the corona wire and the spheres, **a)** on first geometry, **b)** on second geometry.

**Table 2.** Forces along the x-axis of the capacitor of the first geometry.

| Geometry | $F_{px}$ (N/m) | $F_{vx}$ (N/m) | $F_{HTx}$ (N/m) | $F_{ex}$ (N/m) | $F_{Tx}$ (N/m) |
|---|---|---|---|---|---|
| **Corona wire** | -4.264×10$^{-5}$ | -6.496×10$^{-5}$ | -1.076×10$^{-4}$ | 2.887×10$^{-4}$ | 1.811×10$^{-4}$ |
| **Upper sphere** | 2.368×10$^{-4}$ | -3.462×10$^{-4}$ | -1.094×10$^{-4}$ | -1.134×10$^{-1}$ | -1.135×10$^{-1}$ |
| **Lower sphere** | 2.300×10$^{-4}$ | -3.516×10$^{-4}$ | -1.216×10$^{-4}$ | -1.134×10$^{-1}$ | -1.135×10$^{-1}$ |
| **Total force** | 4.242×10$^{-4}$ | -7.628×10$^{-4}$ | -3.386×10$^{-4}$ | -2.265×10$^{-1}$ | -2.268×10$^{-1}$ |

**Table 3.** Forces along the x-axis of the capacitor of the second geometry.

| Geometry | $F_{px}$ (N/m) | $F_{vx}$ (N/m) | $F_{HTx}$ (N/m) | $F_{ex}$ (N/m) | $F_{Tx}$ (N/m) |
|---|---|---|---|---|---|
| **Corona wire** | -6.965×10$^{-5}$ | -8.439×10$^{-5}$ | -1.540×10$^{-4}$ | 2.866×10$^{-4}$ | 1.326×10$^{-4}$ |
| **Ground upper** | 3.867×10$^{-4}$ | -4.358×10$^{-4}$ | -4.910×10$^{-5}$ | 5.040×10$^{-2}$ | 5.035×10$^{-2}$ |
| **Ground lower** | 3.868×10$^{-4}$ | -4.354×10$^{-4}$ | -4.860×10$^{-5}$ | 5.070×10$^{-2}$ | 5.065×10$^{-2}$ |
| **Cathode upper** | 4.354×10$^{-5}$ | -2.391×10$^{-4}$ | -1.956×10$^{-4}$ | -2.283×10$^{-1}$ | -2.285×10$^{-1}$ |
| **Cathode lower** | 4.302×10$^{-5}$ | -2.382×10$^{-4}$ | -1.952×10$^{-4}$ | -2.283×10$^{-1}$ | -2.285×10$^{-1}$ |
| **Total force** | 7.904×10$^{-4}$ | -1.433×10$^{-3}$ | -6.425×10$^{-4}$ | -3.552×10$^{-1}$ | -3.559×10$^{-1}$ |

The electrostatic force on the corona wire is very strong but mostly symmetric (figure 7). Nevertheless, there is a slight asymmetry in the positive ion cloud distribution around the corona wire towards the ground electrode (figures 5 and 6). In this way the electrostatic force on the corona wire will be small and directed towards the spheres, due to the positive ion distribution around it. Therefore the main electrostatic force is concentrated on the grounded and negatively charged spheres (figure 7).

## 4. Conclusion

The present configurations generate a considerable ion wind "beam" between the electrodes (distanced by 1.5 cm) with a velocity of approximately 4-5 m/s which can be useful to substitute mechanical small wind fans with the advantage of no moving mechanical parts (or for other applications as mentioned in the introduction).

As expected, the ion density (figures 5 and 6) around the corona wire is asymmetric causing the force on the corona wire to be to the right. One can also notice that the ion density prolongs much more to the right than to the left of the wire, certainly because of the pull from the grounds and cathodes. The ion density goes to zero right after the grounds end (figure 5) or after the cathodes end (figure 6) because the remaining positive ions are attracted to the ground/cathode and neutralized there.

We can observe that the module of the force on the ground electrodes, in the second geometry, decreases to half (in relation to the grounds in the first geometry) and its direction is inverted towards the cathodes; instead of being towards the corona wire. This is certainly due to the fact that the ground is being pulled into two different directions (to the right and left) that are in competition. In the present conditions, both due to its voltage and distance relative to the ground, the cathode electrodes exert the higher force on the ground and therefore the ground will be pulled to the right. Looking at Table 3 we can see that the second geometry is the best performing setup simulated, certainly demonstrating that charging the extra spheres negatively increases the electrostatic force on them by the positive volume ion cloud, despite the fact that the force on the ground electrodes is in a counter-productive direction.

The first initial geometry [11] mentioned in the introduction developed a force of -0.182 N/m and a maximum ion wind velocity of 3.426 m/s. The first geometry studied here manages to slightly increase these values to -0.227 N/m and 3.898 m/s. But the second geometry really makes a difference with a force of -0.356 N/m it manages to practically duplicate the generated force, and with a maximum ion wind velocity at 4.895 m/s it increases the ion wind velocity by 1.5 m/s.

The second considered initial geometry [12] developed at best a force of -0.327 N/m and a maximum ion wind velocity of 3.915 m/s (ground with 6 cm wide for 2 cm height dimension). Comparing these values with the second geometry studied here we see that the force increases slightly by 0.03 N/m and the ion wind velocity increases by 1 m/s. Since we are now using 0.5 cm diameter spheres (or section of a cylinder) for the ground and cathode (instead of a ground with 6 cm/2 cm that has a bigger frontal area), the present design is much more practical, compact, light, cost effective (uses less material and cheaper to produce shapes) and has a higher performance.

## Acknowledgements

The author gratefully thanks to Mário Lino da Silva for the permission to use his computer with 32 gigabytes of RAM and two quad-core processors, without which this work would not have been possible.